\documentclass[9pt, A4]{article}
\usepackage[left=2.5cm, right=2.5cm, top=1.785cm, bottom=2.0cm]{geometry}
\usepackage{amsmath}
\usepackage[utf8]{inputenc}
\usepackage{graphicx} 
\usepackage[usenames,dvipsnames]{xcolor}
\usepackage{caption}
\usepackage{color}
\usepackage{bm}
\usepackage{authblk}

\title{Microscopic insights into the failure of elastic double networks}
\author[1,a]{Justin Tauber}
\author[1,b]{Simone Dussi}
\author[1,c]{Jasper van der Gucht}
\affil[1]{Physical Chemistry and Soft Matter, Wageningen University, Stippeneng 4, 6708 WE, Wageningen, the Netherlands}
\affil[ ]{}
\affil[a]{\textit {justin.tauber@wur.nl}}
\affil[b]{\textit {simone.dussi@wur.nl}}
\affil[c]{\textit {jasper.vandergucht@wur.nl}}
\date{}                     
\begin{document}

\maketitle

\begin{abstract}
The toughness  of a polymer material can increase significantly if two networks are combined into one material. This toughening effect is a consequence of a transition from a brittle to a ductile failure response. Although this transition and the accompanying toughening effect have been demonstrated in hydrogels first, the concept has been proven effective in elastomers and in macroscopic composites as well. This suggests that the transition is not caused by a specific molecular architecture, but rather by a general physical principle related to the mechanical interplay between two interpenetrating networks. Here we employ theory and computer simulations, inspired by this general principle, to investigate how disorder controls the brittle-to-ductile transition both at the macroscopic and the microscopic level. A random spring network model featuring two different spring types, enables us to study the joined effect of initial disorder and network-induced stress heterogeneity on this transition. We reveal that a mechanical force balance gives a good description of the brittle-to-ductile transition. In addition, the inclusion of disorder in the spring model predicts four different failure regimes along the brittle-to-ductile response in agreement with experimental findings. Finally, we show that the network structure can result in stress concentration, diffuse damage and loss of percolation depending on the failure regime. This work thus provides a framework for the design and optimization of double network materials and underlines the importance of network structure in the toughness of polymer materials.
\end{abstract}

\section{Introduction}
\begin{figure}
    \centering
    \includegraphics{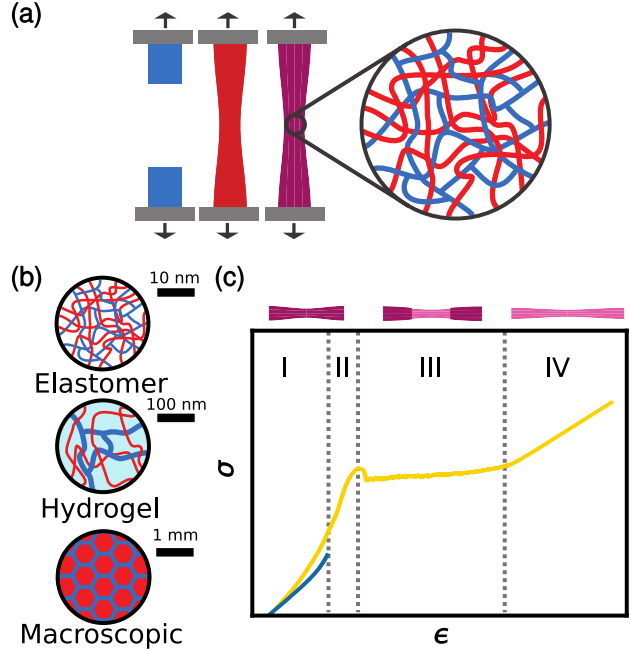}
    \caption{ Double networks across the scale. (a) By combining a sacrificial network that is stiff and weak (blue), with a matrix network that is soft and strong (red), a (molecular) composite both stiff and strong (tough) can be created. (b) Typical double networks are made of elastomers~\cite{Ducrot2014} and hydrogels~\cite{Gong2010}, but they can also be made in the form of macroscopic composites~\cite{Takahashi2018}. (c) Schematic of a brittle (blue) and a ductile (yellow) stress-strain response. }
    \label{fig:Mechanism}
\end{figure}

Polymer networks, such as rubbers and gels, can undergo large deformation without losing elasticity. This property makes them ideal for numerous applications, for example in the biomedical field or in soft robotics. One of the factors that limits their applicability, however, is the brittleness of many polymer networks, which leads to inferior mechanical performance, such as low fracture toughness and strength~\cite{Creton2017}. In recent years, several strategies have been developed to toughen polymer networks by introducing dissipation mechanisms that delay crack propagation. One of the most successful strategies relies on the combination of two different polymer networks into one material, to create a so-called double network. In particular, extremely tough double network hydrogels have been produced by interpenetrating a stiff and weak first network with a soft and extensible second network~\cite{Gong2010,Nakajima2013,Gong2003,Weng2008,Sun2012,Nakajima2012,Xin2013,Nakajima2013a,Chen2014,Yan2017} (Fig.~\ref{fig:Mechanism}(a)). Later, this toughening strategy was also shown to be effective for multi-network elastomers~\cite{Ducrot2014,Millereau2018} or macroscopic composites~\cite{Takahashi2018,King2019}(Fig.~\ref{fig:Mechanism}(b)). This suggests that the underlying principle that leads to toughening is not strongly dependent on molecular details, but governed more generally by an interplay between two mechanically different networks. 

The increase in toughness of double networks is related to a transition from brittle to ductile failure. This brittle-to-ductile transition (BDT) becomes apparent when looking at the stress-strain response of double networks subjected to tensile deformation, as schematically shown in Fig.~\ref{fig:Mechanism}(c). In contrast to single polymer networks that typically fail in a brittle fashion (blue curve), tough double networks fail in a ductile manner with significant softening and plastic deformation (leading to `necking') prior to failure (yellow curve)~\cite{Matsuda2016,Millereau2018}. This macroscopic softening is believed to be due to the progressive breaking of bonds in the stiff network (which is also called the sacrificial network), while the soft matrix formed by the second network stays intact. To obtain tough materials, it is thus crucial that the rupture of the first network occurs without leading to macroscopic cracks in the material. 

Several theoretical models have been proposed to explain the toughness of double networks.  The first of these are due to Tanaka~\cite{Tanaka2007} and Brown~\cite{Brown2007} that describe failure around a defect, such as a notch or a crack. They assumed that a crack tip in the material is surrounded by a region in which the first network breaks by forming microcracks that are stabilized by bridging chains from the second network. Later extensions of these models explicitly accounted for the evolution of damage in the material and were able to describe the hysteresis in the stress-strain response, the progressive softening and the Mullins effect~\cite{Wang2011,Zhao2012,Bacca2017,Vernerey2018,Lavoie2019,Okumura2004}. These phenomenological models assume that the stress field in the material can be described by continuum mechanics, which treats the material as a homogeneous elastic solid. However, it is known that double networks contain large heterogeneities, which play an important role in determining their mechanics~\cite{Kawauchi2009}. To understand how microcracks in the material nucleate and propagate and to find a criterion that predicts the onset of macroscopic failure, a model is needed that takes these heterogeneities into account.

One way to go forward is to include molecular details, for example using molecular dynamics simulations~\cite{Jang2007,Wang2017,Higuchi2018}. However, given the computational costs it is difficult to do this for large system sizes and to explore a large range of parameters. Alternatively, statistical models, such as the fiber bundle model, can give insight into the role of disorder on network failure~\cite{Kovacs2013,Roy2017}. In such models, however, a priori assumptions must be made about the (re)distribution of stress among the different elements, so that they cannot explain the spatial evolution of damage in the materials and its relation to the structure of the networks.

Here, we develop a random double spring network model, which explicitly takes into account disorder and the resulting heterogeneous stress distributions. Rather than making assumptions about the distribution of strain and stress in the material, this distribution emerges naturally from the condition of mechanical equilibrium between the different networks. Because a network is included explicitly, the model provides information on the failure process both at the macroscopic and the microscopic level. We demonstrate how this model predicts a simple criterion for the BDT and we show how the nature of this transition is influenced by disorder and the resulting stress localization. We also compare our network model to a simple 1D multi-spring model, which can be solved analytically. 

\section{\label{sec:Models} Models and Methods}

The BDT is the universal mechanical feature of double network materials. In a brittle material, a small microcrack in the sacrifial network directly leads to macroscopic failure. On the contrary, a ductile material remains intact when microcracks develop. To describe this transition, we first present a simple 1D multi-spring model, which is an extension of a previously used two-spring model~\cite{Ahmed2014, Okumura2004} and which can be solved analytically (Fig.~\ref{fig:ModelOverview}(a)). While heterogeneity can be included in this 1D model, it does not take into account the network structure of the material and therefore cannot give an accurate description of the stress distribution in the material. We therefore consider a more realistic double random spring model that does allow for a heterogeneous stress distribution in the material (Fig.~\ref{fig:ModelOverview}(b)).  

\subsection{ Multi-spring model}

\begin{figure}
    \centering
    \includegraphics{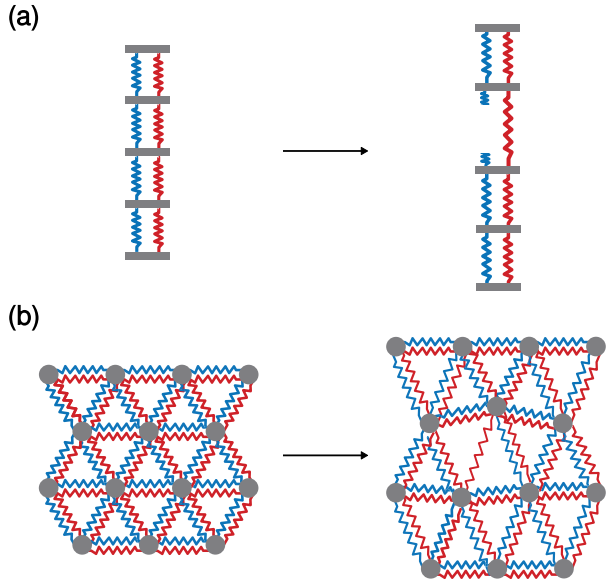}
    \caption{Minimal models for elastic double networks and their reaction to failure. The springs represent elements of the sacrificial network (blue) and matrix network (red). (a) Multi-springs (MS) model. (b) Random Spring Network (RSN) model. Simulations are performed for networks of $50\times50$ nodes.}
    \label{fig:ModelOverview}
\end{figure}

In the multi-spring model (MS), the double network is modeled as a series of elements, each consisting of two parallel springs, one representing the sacrificial network (with spring constant $\mu_S$) and one representing the matrix (with spring constant $\mu_M$) (see Fig.~\ref{fig:ModelOverview}(a)).  Each spring is assumed to be a linear Hookean spring, so that the force acting on it is given by $F_i=\mu_i\Delta l_i$, with $\Delta l_i$ the extension of spring $i$. 
Removal of a sacrificial spring in an element leads to the formation of a microcrack, which is bridged by the remaining matrix spring. As the elements are connected in series, the load on each element is equal. This force balance between the elements implies

\begin{equation}\label{forcebalance}
\sigma=(\mu_M+\mu_S)\Delta l_D = \mu_M\Delta l_M \; ,
\end{equation}

\noindent where $\Delta l_D$ and $\Delta l_M$ represent the extension of the network in the intact double network and in the matrix spanning the microcrack, respectively, and where $\sigma$ denotes the stress. The overall strain of the system $\epsilon$ can be written as

\begin{equation}\label{strain}
\epsilon=\Delta l_D(1-\phi)+\Delta l_M \phi  \; ,
\end{equation}

\noindent where $\phi$ denotes the fraction of broken sacrificial bonds (i.e. the fraction of microcracks). 

Rupture of a bond is assumed to occur instantaneously when the extension of a spring $\Delta l_i$ exceeds its threshold $\lambda_i$. In the absence of disorder, all sacrificial bonds have the same threshold $\lambda_S$ and all matrix bonds have the same threshold $\lambda_M$. Brittle failure after formation of a microcrack occurs when the bridging matrix bond reaches its threshold before the other sacrificial bonds, while ductile failure occurs when the sacrificial bonds reach their threshold first. From equation~\ref{forcebalance}
it immediately follows that the brittle-to-ductile transition occurs when $(\mu_S+\mu_M)\lambda_S=\mu_M\lambda_M$. We can therefore predict which system parameters are required for a system at the BDT. For example, the sacrificial bond threshold $\lambda_S^{*}$ for which the rupture force of a (matrix-reinforced) sacrificial bond is equal to the rupture force of a matrix bond is given by

\begin{equation}\label{eq:lambdaSstar}
    \lambda_S^{*} = \frac{\mu_M\lambda_M}{\mu_S + \mu_M}  \;.
\end{equation}

\noindent More generally, we can define a parameter $\Delta\alpha$ to quantify the distance from the BDT:

\begin{equation}\label{Da}
\Delta\alpha=\frac{\mu_M\lambda_M}{(\mu_M+\mu_S)\lambda_S}-1  \; .
\end{equation}

If $\Delta\alpha=0$ the system is at the BDT. Brittle failure occurs for $\Delta\alpha<0$; in this case only one microcrack is enough to cause global failure so the fraction of broken sacrificial bonds at failure is $\phi_f=1/N$ with $N$ the system size. Ductile failure occurs for $\Delta\alpha>0$, where all sacrificial bonds break before the system fails globally, so $\phi_f=1$. It follows from this analysis that for ductile failure we need $\lambda_S<\lambda_M$, which explains why the creation of tough, ductile networks requires the sacrificial network to be much weaker than the matrix.

For a perfectly homogeneous system, the BDT in this model is an abrupt transition. Experiments, however, show that the descriptors of failure vary in a continuous manner through the BDT~\cite{Matsuda2016}. We therefore introduce disorder by assuming that the thresholds of the sacrificial network vary according to a certain distribution $P(\lambda_S)$. All other parameters are taken the same. In this case, the sacrificial bonds will fail progressively, with the weakest bonds breaking first and the stronger bonds breaking later. At a given strain $\epsilon$, all sacrificial bonds for which $\lambda_S<\Delta l_D$ are broken. In the continuous limit (for $N\gg 1$), the fraction of broken sacrificial bonds can thus be written as

\begin{equation}\label{dist}
\phi(\epsilon)=\int_0^{\Delta l_D} P(\lambda_S)\textrm{d}\lambda_D  \; .
\end{equation}

Here, we will consider a Gaussian threshold distribution with mean $\langle\lambda_S\rangle$ and standard deviation $\delta\lambda$. For this case, it follows from equations \ref{forcebalance} and \ref{dist} that at the moment of macroscopic failure (where $\Delta l_M=\lambda_M$), the fraction of broken sacrificial bonds is given by

\begin{equation}\label{dist2}
\phi_f=\frac{1}{2}\left[1+\textrm{erf}\left(\Delta\tilde{\alpha}\right)\right]  \; ,
\end{equation}

\noindent with 

\begin{equation}\label{Datilde}
\Delta\tilde{\alpha}=\frac{\Delta\alpha\langle\lambda_S\rangle}{\delta\lambda\sqrt{2}}  \; ,
\end{equation}

\noindent a normalized parameter to quantify the distance to the BDT, which takes into account the disorder in the thresholds. The fraction of broken sacrificial bonds thus increases gradually from 0 to 1 along the BDT and the parameter $\Delta\tilde{\alpha}$ sets the steepness of this transition. With increasing disorder, the transition becomes more gradual. The strain at break follows from equations \ref{forcebalance} and \ref{strain}:

\begin{equation} \label{eq:failureStrain}
\epsilon_f=\lambda_M\left(\frac{\mu_M+\phi_f\mu_S}{\mu_M+\mu_S}\right)  \;.
\end{equation}

In the brittle regime ($\Delta\tilde{\alpha}\ll-1$, $\phi_f\approx0$) we find $\epsilon_f=\mu_M\lambda_M/(\mu_M+\mu_S)$, while in the ductile regime ($\Delta\tilde{\alpha}\gg1$, $\phi_f\approx1$) we have $\epsilon_f=\lambda_M$. 

\subsection{Random Spring Network Model}
Being a 1D model, the multi-spring model cannot account for stress heterogeneity and the resulting localization of stresses. We therefore consider a random spring network model (RSN) as shown in Fig.~\ref{fig:ModelOverview}(b) composed of $L \times L$ nodes arranged on a triangular lattice whose nearest neighbours are connected by the same element we introduced for the multi-spring model ($L=50$ for all RSN simulations). Now, upon applying an external load, stress concentration is possible due to the topological restrictions imposed by the network structure.

Following the experimentally found guidelines for making tough double networks~\cite{Gong2010} we consider materials in which the sacrificial network is stiff and weak, while the matrix is soft and strong. This means that the elastic constant of sacrificial springs $\mu_S$ is always higher than the one of the matrix springs $\mu_M$. We therefore vary the ratio of the stiffnesses such that $0<\mu_M/\mu_S<1$. To reflect the asymmetry in network strengths, we fix $\lambda_M=4.0$ and vary $\lambda_S$ such that $\lambda_S<\lambda_M$. In all cases, we keep the parameters of all matrix springs the same. To implement disorder at the bond-level, the failure thresholds $\lambda_S$ are picked from a Gaussian distribution as in the MS model. Nodes at the bottom and top are fixed in the $y$-direction, but they can slide along the $x$-direction. Along the $x$-axis periodic boundary conditions are implemented. 

We characterize the mechanical response of the network by applying an extensional strain $\epsilon$ along the y-axis in small steps of 0.1\% strain. We consider quasi-static loading and assume that the networks remain in mechanical equilibrium at each step, i.e. it settles in its minimum energy state. Therefore, after every step, the total energy of the network is minimized by displacing the nodes using the FIRE algorithm~\cite{Bitzek2006} with a tolerance of $F_{rms}\leq1\cdot10^{-5}$. Here, $F_{rms}$ is the maximum root mean squared force allowed in the system~\cite{Bitzek2006}. To simulate failure, we break all bonds which exceed their failure threshold $\lambda_i$ consecutively. After the failure of every single bond the energy of the system is minimized. Once all overstretched bonds are broken, the next strain step is determined according to the bond that is closest to its failure threshold with a minimum step size of 0.001\% strain. 

For every strain in the simulations the virial stress is calculated from the forces exerted by the springs on the nodes. As the system is elongated along the $y$-direction, we consider the $yy$-component of the virial tensor as a measure for the stress $\sigma$. We define the softening strain $\epsilon_{soft}$ as the strain where the stiffness drops below the initial stiffness, and the failure strain $\epsilon_f$ and failure stress $\sigma_f$ as the strain and stress for which the system becomes disconnected along the y-direction. For every data point 50 simulations are performed, errorbars represent standard deviation.

\section{\label{sec:ResultsDiscussion} Results \& discussion}

\begin{figure}
    \centering
    \includegraphics{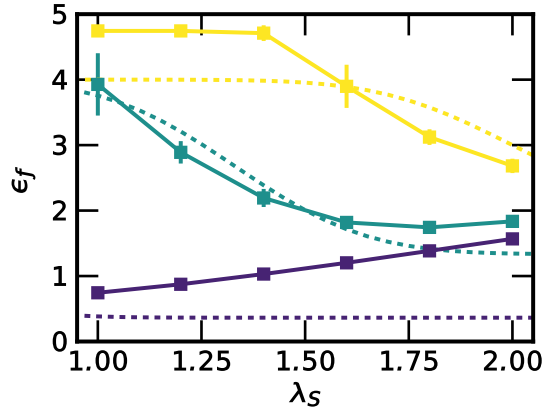}
    \caption{Failure strain, $\epsilon_f$, versus the average failure threshold $\langle \lambda_S \rangle$ in systems with $\delta\lambda=0.250$. Results are shown for the RSN model (solid lines) and the MS model (dashed lines) for stiffness ratios $\mu_M/\mu_S =$ 0.10 (purple), 0.50 (green) and 1.00 (yellow)).}
    \label{fig:FailureStrain}
\end{figure}

\begin{figure}
    \centering
    \includegraphics{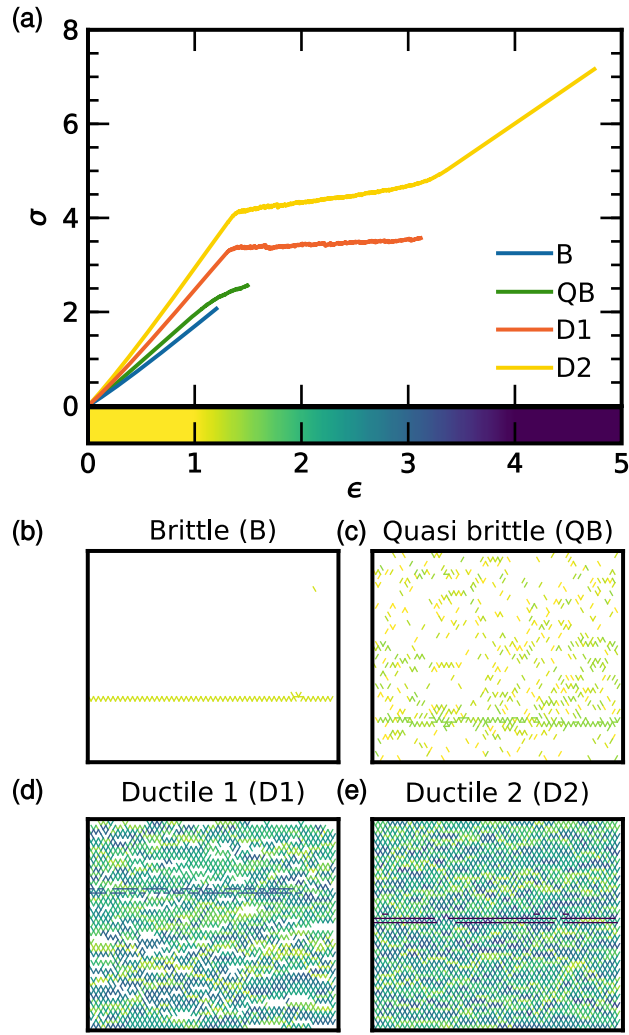}
    \caption{Typical failure response of double networks obtained via the RSN model. (a) The four distinct stress-strain responses: brittle (B), quasibrittle (QB), the first ductile regime (D1), and the second ductile regime (D2). (b)-(e) Corresponding failure patterns. Every line represents a broken sacrificial bond, color-coded according to the strain at which it failed (see color bar above).}
    \label{fig:StressStrain}
\end{figure}

\subsection{\label{sec:FailureRegimes} Failure regimes of double networks}

The aim of this paper is to systematically explore the effect of disorder and stress heterogeneity on the failure of double networks, focusing especially on the role of the network. To do this in a systematic fashion, we will extensively compare the results of the random spring network (RSN) model with the 1D multi-spring (MS) model lacking any network structure.

We start our comparison by checking a prediction from the MS model for the dependence of the failure strain $\epsilon_f$, described in equation~\ref{eq:failureStrain}, on the sacrificial bond properties $\lambda_S$ and $\mu_M/\mu_S$ in the presence of disorder (Fig.~\ref{fig:FailureStrain}). Although it is clear that both the MS model and the RSN model follow the same trend, the overall trend is not immediately obvious. Intuitively, it is expected that the failure strain should increase with an increase in $\lambda_S$. However, only at the lowest stiffness ratio, we see that the $\epsilon_f$ increases with $\lambda_S$. At high stiffness ratios, this trend is reversed.
This transition is a consequence of the brittle-to-ductile transition (BDT). From equation~\ref{Da} it can be deduced that an increase in $\lambda_S$ brings the system from the ductile into the brittle regime. The transition is not observed at the lowest stiffness ratio, because all those systems are already in the brittle regime.
Similar trends can be found for the material strength and the work of extension (see Supplemental Material~\cite{supp}).

These results show that we can tune the failure response based on the properties of the individual sacrificial bonds. Taking a broader view at the entire stress-strain response of the RSN, we can differentiate between four distinct responses, shown in Fig.~\ref{fig:StressStrain}(a). There are two brittle responses, brittle (B) and quasibrittle (QB), that can be distinguished by the (slight) softening that occurs before macroscopic failure in the QB case. There are also two ductile responses (D1 and D2) that both show a clear plateau in the stress-strain response after softening. In case of D1, macroscopic failure occurs within this plateau. For D2, a second increase in the stress precedes final failure. These same four responses are also found in the MS model (see Supplemental Material~\cite{supp}).

By contrast, at the microscopic level the failure behaviour of the RSN is distinct from the MS model. In Fig.~\ref{fig:StressStrain}(b-e), we plot the broken bonds of the sacrificial network color-coded according to the strain at which they break. We clearly observe that the number of broken sacrificial bonds increases from B to D2 type of fracture. In B, all broken bonds are part of a single crack at the location where macroscopic failure occurs. A similar pattern is observed for QB, where, however, more bonds fail homogeneously throughout the sample before the final crack appears. It is the failure of these bonds that causes the softening observed in the QB stress-strain response. The ductile regimes are characterized by many bonds breaking simultaneously. Such avalanches occur over a large strain range, from the moment of softening until macroscopic failure. In case of D2 fracture, all the non-horizontal bonds of the sacrificial network, which are the elastically active bonds, are broken. In the MS model we can not differentiate between failure patterns, because the load is homogeneous throughout the system, so that stress can not concentrate and damage will always be homogeneous.

We note that both the RSN model and the MS model capture the four different mechanical responses observed in experiments on elastomers~\cite{Millereau2018}, hydrogels~\cite{Gong2010} and macroscopic composites~\cite{Takahashi2018,King2019}. The clear difference in fraction of broken sacrificial bonds between (quasi)brittle and ductile responses is also well established experimentally~\cite{Matsuda2016,Millereau2018}. Furthermore, the softening response, which is irreversible both in our model and in the experiments, is attributed to early damage of bonds in the sacrificial network~\cite{Vernerey2018} which has been shown to occur homogeneously throughout the network well before macroscopic failure~\cite{Millereau2018}. Finally, also a plateau in the stress response is typically observed in experiments after the yielding of the material, often accompanied by a necking region where only sacrificial bonds break~\cite{Millereau2018}. We conclude that both the RSN and MS model represent a minimal yet insightful model to study fracture in elastic double networks both at the macroscopic and microscopic level and therefore we can proceed to a more detailed analysis.

\subsection{\label{sec:MacroResponse} Macroscopic characterization of the failure regimes}

\begin{figure*}
    \centering
    \includegraphics{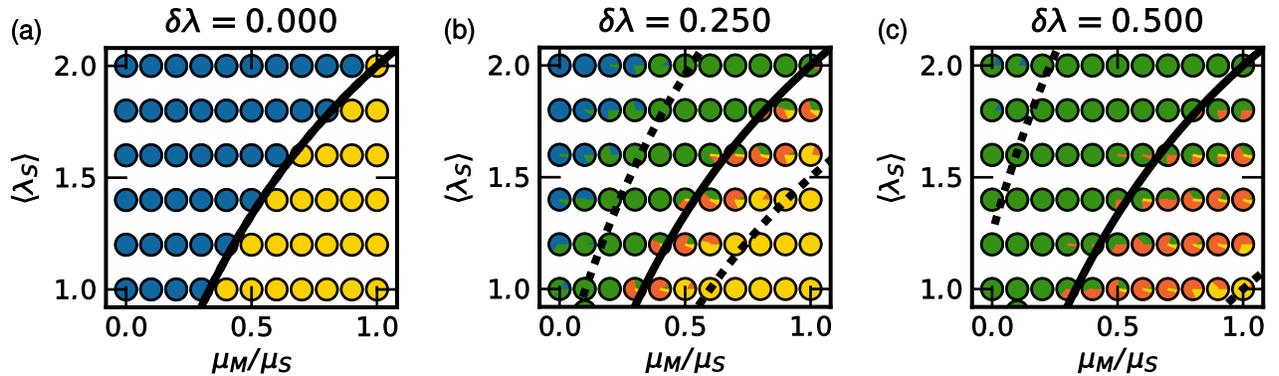}
    \caption{Macroscopic mechanical phase diagrams in the $\left(\mu_M/\mu_S,\lambda_S\right)$-plane. Each panel corresponds to a different value of disorder $\delta\lambda$. Symbols are pie-charts color-coded according to the frequency of the four fracture regimes (B (blue), QB (green), D1 (orange), D2 (yellow)) observed from 50 independent simulations. The black line indicates the BDT according to the MS model. The dashed lines indicate the B-to-QB transition ($\Delta\tilde{\alpha}=-2.5$) and the D1-to-D2 transition ($\Delta\tilde{\alpha}=2.0$) in the MS model (see Supplemental Material~\cite{supp}).}
    \label{fig:PhaseDiagram}
\end{figure*} 

The results in Fig.~\ref{fig:FailureStrain} clearly demonstrate the dependence of the failure response on the properties of the sacrificial bonds: $\langle \lambda_S \rangle$, and $\mu_M/\mu_S$. To study the joined effect of these properties together with the disorder in the failure threshold ($\delta\lambda$), we construct phase diagrams of the failure response (Fig.~\ref{fig:PhaseDiagram}) based on the four different regimes described above (see Supplemental Material~\cite{supp}). For each point in the diagram we use a pie-chart symbol indicating the frequency of the four different regimes observed over 50 independent simulations. We first focus on the case with no disorder $\delta\lambda=0$, shown in Fig.~\ref{fig:PhaseDiagram}(a), and observe that both $\mu_M/\mu_S$ and $\lambda_S$ control the sharp transition from the BDT. Also, the BDT of the RSN model coincides with the BDT of the MS model (Eq.~\ref{Da}) as indicated by the black line. 

Next, we consider the cases with disorder, plotted in Fig.~\ref{fig:PhaseDiagram} (b) and (c), and observe the appearance of the intermediate regimes QB and D1. These two regimes appear around the BDT and upon increasing disorder $\delta \lambda$ progressively span the entire explored parameter space, which is a clear indication that disorder controls the position of these transitions. In addition, the transitions between the regimes are less sharp when the disorder is larger, as testified by the less homogeneously colored pie-chart symbols. Indeed, the mechanical responses of networks with large $\delta \lambda$ depend on the exact realization of the network (exact distribution of threshold values and their spatial organization), even when having the same $\mu_M/\mu_S$ and $\left<\lambda_S\right>$, especially around boundaries between different regimes.
A similar dependence of the location of the B-to-QB and D1-to-D2 transitions on disorder is also predicted by the MS model (dashed lines, see Supplemental Material for further information~\cite{supp}). At low disorder, the location of the boundaries corresponds well between the models, but at high disorder the boundaries move away from each other. This comparison reveals that the type of load sharing (equal load sharing in the MS model or network controlled load sharing in the RSN model) is not essential for the occurrence of any of the four regimes. Nevertheless, the stress redistribution via the network does affect how far from the BDT these regimes occur. 
Furthermore, disorder clearly plays an important role also in the experimental systems since QB and D1 responses are typically observed experimentally and we show that these occur only in presence of disorder. Our results confirm that knowing the average value of the bond strengths might not be enough to predict the failure regime, in agreement with recent phenomenological models aiming to capture softening and the Mullins effect in double networks~\cite{Vernerey2018}.

\subsection{\label{sec:MicroResponse} Failure at the microscopic level}
In the previous section, we found that the macroscopic force balance reasonably captures the transition from (quasi)brittle to ductile fracture in the RSN model and that disorder influences this transition. Furthermore, we have seen in Fig.~\ref{fig:StressStrain}(b)-(e) that as long as some disorder is present distinct failure patterns develop in the RSN that suggest a tight link between the macroscopic and microscopic failure process. In this section, we explore several indicators of microscopic failure in disordered RSN systems, with a special focus on stress heterogeneity, and we consider to what extent the force balance influences the microscopic failure processes in double networks.

\subsubsection{The fraction of broken bonds}

\begin{figure}
    \centering
    \includegraphics{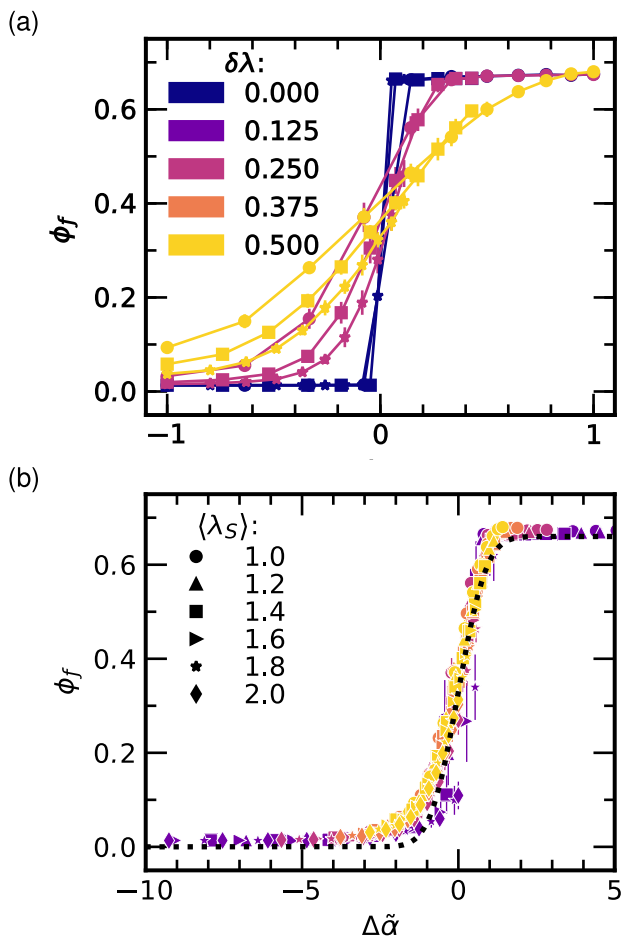}
    \caption{Rescaling of the fraction of broken bonds $\phi_f$. (a) The fraction of total broken bonds $\phi_{f}$ obtained in simulations of RSN is plotted against $\Delta\alpha$, which quantifies the distance from the BDT predicted by the equal load sharing models. (b) $\phi_{f}$ from RSN versus $\Delta\tilde{\alpha}$. The dashed black line indicates the theoretical prediction for $\phi_f$ according to the MS model (equation \ref{Da}). In the RSN model only the vertically aligned bonds are expected to break, therefore the prediction of the MS model is multiplied by a factor $2/3$.} The symbols correspond to the $\lambda_{S}$ value (see legend panel (b)) color-coded according to the amount of disorder $\delta\lambda$ (see legend panel (a)). Error bars indicate standard deviation based on 50 configurations.
    \label{fig:BrokenBondRescaling}
\end{figure}

The first parameter we explore is the fraction of broken sacrificial bonds at final failure $\phi_f$, as we have a direct prediction from the MS model. In Fig.~\ref{fig:BrokenBondRescaling}(a), we plot $\phi_{f}$ as a function of $\Delta\alpha$, the distance from the BDT as introduced in equation \ref{Da} for various amount of disorder $\delta\lambda$. Clearly, we see a strong increase in the number of broken bonds upon going from the brittle to the ductile regime. For $\delta\lambda=0$ we observe that all data points collapse on a step function exactly at the BDT, $\Delta\alpha=0$. When disorder is introduced, the transition becomes more gradual, corresponding to the shift of the B-to-QB and D1-to-D2 transition in Fig.~\ref{fig:PhaseDiagram}. We can understand these effects by pointing out that both weak and strong bonds are introduced due to the spread in sacrificial bond strength. For $\Delta\alpha<0$, $\phi_{f}$ is increased by the disorder due to the presence of weak springs that break before brittle failure occurs. Similarly, for $\Delta\alpha>0$, $\phi_{f}$ decreases due to disorder that introduces bonds that are too strong to break before the matrix fails. 

According to the prediction for the MS model in equation~\ref{dist2} we can collapse all this data on one master curve for all values of $\lambda_S$, $\mu_M/\mu_S$ and $\delta\lambda$, if we consider the rescaled distance to the BDT: $\Delta\tilde{\alpha}$. Indeed we obtain a good collapse for $\phi_{f}$, as shown in Fig.~\ref{fig:BrokenBondRescaling}(b). However, there is a significant increase in $\phi_f$ in the brittle regime relative to the prediction of the MS model (dashed line) (See Supplemental Material \cite{supp}). This could be an indication that the load sharing via the network enhances the failure of sacrificial bonds, however we can not exclude that the discrepancy is due to a finite size effect. 

\subsubsection{\label{sec:stressConc} Stress concentration in the spring network model}

\begin{figure*}
    \centering
    \includegraphics{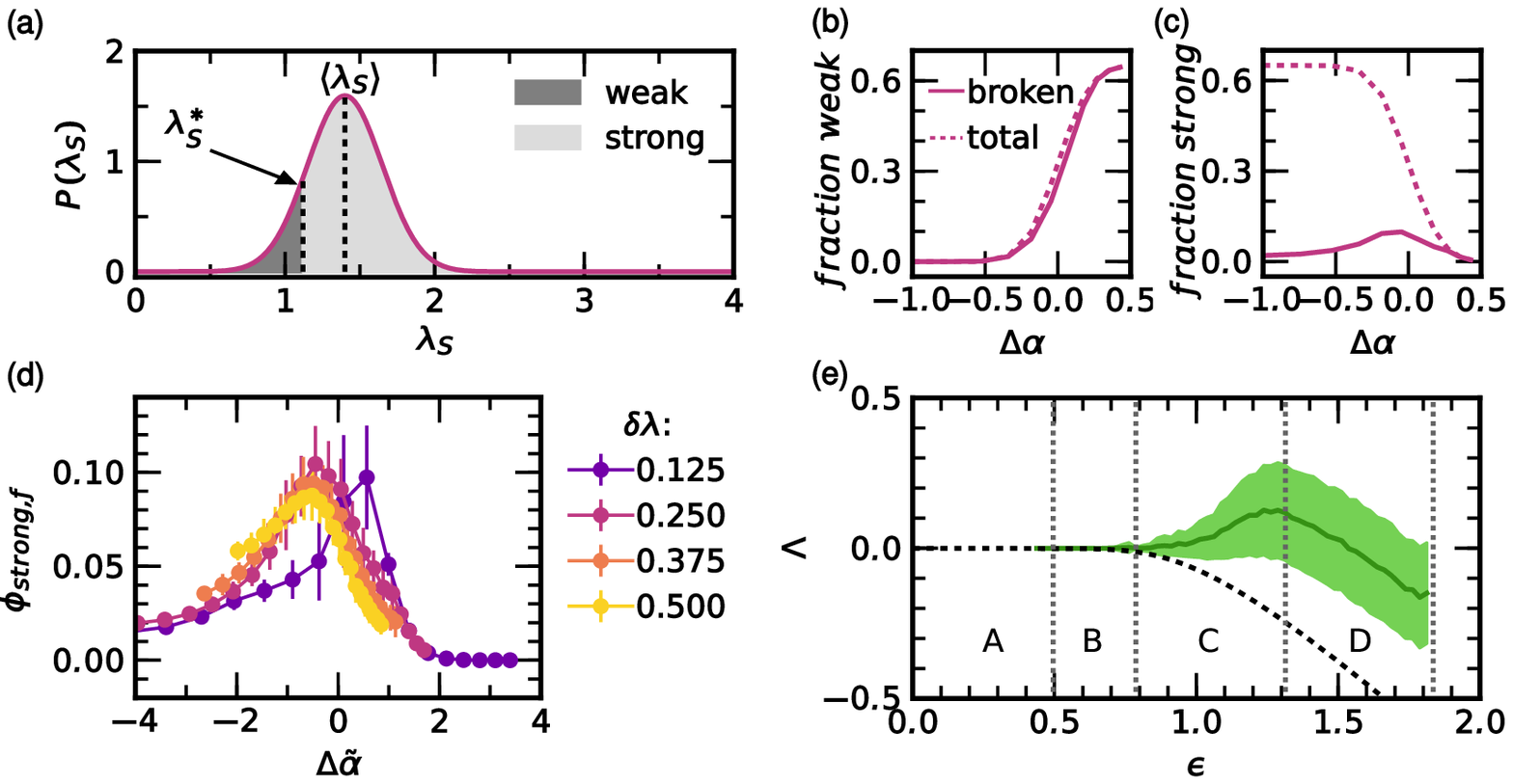}
    \caption{Development of stress heterogeneity and stress concentration during deformation. In all subpanels  $\left<\lambda_S\right>=1.4$. (a) Distribution of $\lambda_S$ for a system with $\left<\lambda_S\right>=1.4$ and $\Delta\tilde{\alpha}=-0.7$. At $\lambda^{*}_{S}$ the force required to break a sacrificial bond is equal to the force required to break a matrix bond. Thus, if $\lambda_S<\lambda^{*}_{S}$ a sacrificial bond is weaker than the matrix bonds (dark grey) and if $\lambda_S>\lambda^{*}_{S}$ a sacrificial bond is stronger (light grey). (b) Comparison of total fraction of weak bonds that align with the deformation field and the fraction of weak bonds $\phi_{weak,f}$ that failed during deformation versus $\Delta\alpha$. (c) Similar comparison for the total amount of strong bonds and broken strong bonds $\phi_{strong,f}$. (d) $\phi_{strong,f}$ versus $\Delta\tilde{\alpha}$ for different $\delta\lambda$. Error bars indicate standard deviation. (e) Development of $\Lambda$, defined as the difference between the rupture threshold $\lambda_{S,fail}$ of the failing sacrificial bonds and $dl_{aff}$, the extension of the bonds if the strain field were homogeneous throughout the system, as a function of strain for a network with $\Delta\tilde{\alpha}=-0.7$, $\langle\lambda_S\rangle=1.4$ and $\delta\lambda=0.250$. The four intervals are described in the text. The average (line) and standard deviation (shaded area) are calculated from a histogram of $\lambda_{S,fail}-dl_{aff}$ over 50 simulations accumulated over bins of $2.5\%$ width. The dashed black line indicates the prediction for $\Lambda$ according to the MS model.}
    \label{fig:StressConc}
\end{figure*}

In the MS model it is assumed that stress is distributed homogeneously throughout the entire system, even after failure of sacrificial bonds. Hence, stress can never concentrate within the MS model. In the RSN model stress concentration is possible and therefore it can be used to study the nucleation and propagation of failure in double networks. It is fascinating that this fundamental difference between the MS model and the RSN model does not result in big differences in macroscopic failure or the fraction of broken sacrificial bonds. 

Interestingly, a second look at the type of sacrificial bonds that break does reveal a striking difference between the MS model and the RSN model caused by the concentration of stress. By construction, in the MS model sacrificial bonds break in the order of their strength, from weak to strong, until a certain threshold $\lambda_{S}^{*}$ (equation~\ref{eq:lambdaSstar}) is reached and macroscopic failure occurs. At this threshold, $\lambda_{S}^{*}$, the force required to break a (matrix-reinforced) sacrificial bond is equal to the force required to break a matrix bond. How $\lambda_{S}^{*}$ influences the failure of bonds is further illustrated in Fig.~\ref{fig:StressConc}(a) for a system below the BDT ($\Delta\tilde{\alpha}=-0.7$, $\mu_M=0.4$ and $\langle\lambda_S\rangle=1.4$). For this system $\lambda_{S}^{*}=1.14$, so in the MS model only weak bonds with $\lambda_S<1.14$ would fail. However, in the RSN model we see that sacrificial bonds stronger than $\lambda_{S}^{*}$ also break (see Fig.~\ref{fig:StressConc}(b) and (c)). In particular, in Fig.~\ref{fig:StressConc}(d) we plot the fraction of broken strong sacrificial bonds $\phi_{strong,f}$ (i.e. bonds with $\lambda_S > \lambda^{*}_{S}$ that break) as a function of the distance from the BDT $\Delta\tilde{\alpha}$ for various amounts of disorder $\delta \lambda$. At low $\delta\lambda$ and far below the BDT, failure of strong sacrificial bonds is required in order to achieve macroscopic failure, since there are just not enough weak sacrificial bonds (Fig.~\ref{fig:StressConc}(b)). However, upon approaching the transition at $\Delta\tilde{\alpha}=0$, $\phi_{strong,f}$ grows, even though a decrease could be expected based on the total concentration of strong sacrificial bonds (Fig.~\ref{fig:StressConc}(c)). Interestingly, we observe a maximum in $\phi_{strong,f}$ before fully entering the ductile regime $\Delta\tilde{\alpha}>0$ where mostly weak sacrificial bonds control the fracture process. Strong bonds can only break if the force they carry exceeds the maximum force that is expected based on the macroscopic force balance. Thus, we conclude that the network structure plays a crucial role in (re)distributing the load during deformation, giving rise to stress concentration. The development of stress heterogeneity is tightly bound to the disorder that is present in the initial structure. We expect that by including more (structural) disorder in the initial network (e.g.inhomogeneities in the connectivity of the nodes, a disordered spatial distribution of the nodes, or a distribution in the stiffness of the bonds) the stress heterogeneity would become even larger~\cite{Dussi2020}. As a consequence, we would expect significant broadening of the QB and D1 phase with respect to the prediction from the MS model. 

Because in the RSN the initial structure is homogeneous, stress concentration does not occur from the onset of deformation, but develops during deformation as bonds are ruptured. We can quantify this development via the difference between the failure threshold of a broken sacrificial bond $\lambda_{S,fail}$, a measure for the actual local extension, and the extension estimated from the global strain assuming all deformations are affine, $dl_{aff}$. We therefore define $\Lambda=\lambda_{S,fail}-dl_{aff}$. To compare with the MS model, it must be noted that in this model the stress is homogeneous, but the strain field is not. In fact, the microcracks take up additional strain to compensate for the absence of the sacrificial network. This means that the deformation of the remaining sacrificial bonds is less than the affine strain, so that $\Lambda$ is always smaller that 1 in the MS model (See dashed line in Fig.~\ref{fig:StressConc}(e)). If at any time $\Lambda>0$, stress concentration must therefore be present.

This analysis reveals different modes of stress concentration during the failure of a RSN system as shown in in Fig.~\ref{fig:StressConc}(e). At first nothing breaks, this regime coincides with the reversibly elastic regime found at the macroscopic level (regime A). Then, sacrificial bonds start to break exactly as predicted by the affine deformation. Here both the average stress concentration, $\langle\Lambda\rangle$, and its variance are zero (regime B). Upon further increasing the deformation, the variance of $\Lambda$ starts to grow indicating stress heterogeneity in the system, quickly followed by an increase of $\langle\Lambda\rangle$ indicating stress concentration (regime C). As a result strong sacrificial bonds also start to break and failure of weak sacrificial bonds is postponed or even prevented (see Supplemental Material~\cite{supp}). Finally we arrive at a peak in $\langle\Lambda\rangle$, after which the $\langle\Lambda\rangle$ decreases until macroscopic failure is reached (regime D). We attribute this decrease to structural relaxation of the sacrificial network via the formation of microcracks, leading to loss of rigidity and eventually loss of percolation in the sacrificial network. It must be noted that the variance in $\langle\Lambda\rangle$ remains constant, indicating that heterogeneity in stress is present until the final failure event.

\subsubsection{Crack development and stress delocalization}

\begin{figure*}
    \centering
    \includegraphics{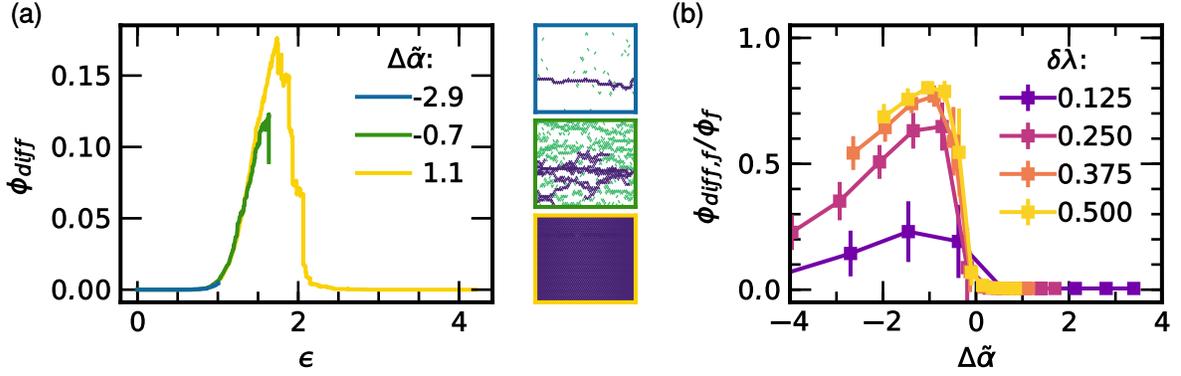}
    \caption{ Evolution of diffuse damage. (a) Fraction of broken sacrificial bonds not belonging to the largest crack ($\phi_{diff}$) as a function of strain $\epsilon$ for different combinations of parameters below ($\Delta\tilde{\alpha}<0$) and above ($\Delta\tilde{\alpha}>0$) the BDT. Corresponding failure patterns in the sacrificial network are also shown, with purple bonds belonging to the largest crack and green bonds to diffuse damage. (b) Normalized diffuse damage at final failure $\phi_{diff,f}$
    as a function of $\Delta\tilde{\alpha}$ for different disorder $\delta\lambda$. Points are averages and error bars indicate standard deviation calculated over 50 configurations. For all curves $\left<\lambda_S\right>=1.4$.}
    \label{fig:Diffuse}
\end{figure*}

The development of stress heterogeneity and stress concentration during deformation, is the result of microscopic damage evolution in the sacrificial network: the nucleation and propagation of cracks. We also suggested that the development of stress concentration is tightly linked to the disorder in the initial, undeformed, system. How crack nucleation and propagation are influenced by both initial disorder and stress heterogeneity during deformation is not trivial. We have already seen that in the RSN model not all sacrificial bonds that break are part of the largest crack (Fig.~\ref{fig:StressStrain}). Thus, it seems that stress can be delocalized away from the crack tip of the largest crack throughout the system via diffuse damage. In this section, we quantify the evolution of (micro)cracks in the sacrificial network, focusing on diffuse damage and its coupling with stress delocalization. 

In Fig.~\ref{fig:Diffuse}(a), we plot the diffuse damage, quantified by the fraction of broken sacrificial bonds (with respect to the total number of initial sacrificial bonds) that are not part of the largest crack $\phi_{diff}$ as a function of strain $\epsilon$ for three representative systems below, around and above the BDT (see Supplemental Material~\cite{supp}). In the brittle case ($\Delta\tilde{\alpha}=-2.9$), $\phi_{diff}$ barely increases and fracture occurs at smaller strain. Closer to the BDT ($\Delta\tilde{\alpha}=-0.7$), the diffuse damage significantly increases, reaches a maximum, and just before final failure slightly decreases. This decrease in the curve indicates that microcracks merge into the largest crack that eventually breaks the system (Fig.~\ref{fig:Diffuse}(a)). Beyond the BDT ($\Delta\tilde{\alpha}=1.1$), the diffuse damage reaches a higher maximum of more than 15\% of the total amount of sacrificial bonds before a significant decrease that extends for a large strain interval. Finally, $\phi_{diff}$ reaches zero, indicating that all microcracks are now merged in a single large crack or damage zone, well before the final failure and all subsequent failing sacrificial bonds join this large crack. 

With respect to the global stress-strain response, diffuse failure is expected to be most influential before the BDT. In the ductile regime, broken bonds will eventually join to form one large damage zone. However, in the brittle regime, many bonds can break in a diffuse fashion without being part of the largest crack, potentially diminishing stress concentration. We therefore focus on the final ratio of the broken sacrificial bonds belonging to the diffuse damage $\phi_{diff,f}$ as an indicator of stress delocalization and plot it versus $\Delta\tilde{\alpha}$ for various amounts of disorder in Fig.~\ref{fig:Diffuse}(b). We observe a non-monotonic behavior with a peak slightly before the BDT. At the BDT $\phi_{diff,f}$ drops drastically as all bonds eventually join the largest crack. We conclude that just before the BDT, the diffuse damage is maximum and therefore hypothesize that the crack nucleation is delayed the most in this region. A consequence of this delayed nucleation is the widening and branching of the largest crack as shown in Fig.~\ref{fig:Diffuse} (a), snapshot in the middle. The widening of the damage zone in the sacrificial network around a defect is often used as an explanation for the enhanced fracture toughness in double network materials\cite{Brown2007,Gong2010}. Although our simulations do not measure the fracture toughness directly, our results suggest that widening of the damage zone and therefore fracture toughness is maximal just before the BDT.

\subsection{\label{sec:MatProp} Relating microscopic events to the macroscopic failure regimes}

\begin{figure}
    \centering
    \includegraphics{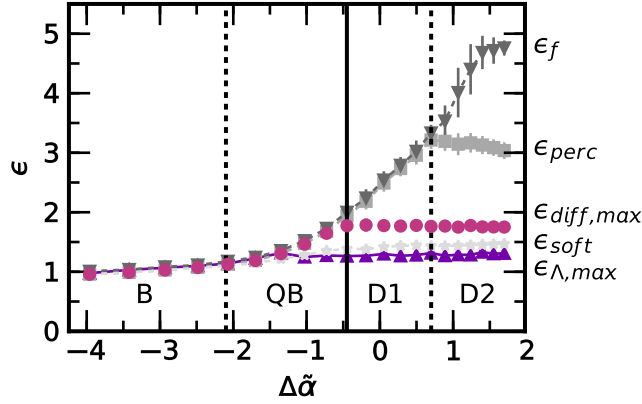}
    \caption{Characteristic strains corresponding to microscopic and macroscopic events as a function of $\Delta\tilde{\alpha}$ ($\langle\lambda_S\rangle=1.4$ and $\delta\lambda=0.250$). $\epsilon_{\Lambda,max}$ (purple triangles) is the strain at maximum stress concentration (see Fig.~\ref{fig:StressConc}(e)). $\epsilon_{diff,max}$ (pink circles) is the strain at which the diffuse damage is maximal (see Fig.~\ref{fig:Diffuse}(a)). $\epsilon_{perc}$ gray squares) is the strain at which percolation in the sacrificial network is lost. $\epsilon_{soft}$ (light gray downward triangles) indicates the onset of softening in the stress-strain curve (see Fig.~\ref{fig:StressStrain}(a) and Supplemental Material \cite{supp}). $\epsilon_f$ (dark gray stars) corresponds to final fracture. The vertical lines indicate the transitions between the failure regimes in the RSN estimated from the phase diagram in Fig.~\ref{fig:PhaseDiagram}(b).}
    \label{fig:MacroToMicro}
\end{figure}

The network and the heterogeneous stress distribution emerging upon deformation, clearly have a huge impact on the microscopic failure mechanism. We have seen that which microscopic failure mechanisms occur is controlled by the distance from the BDT. In addition, we have observed that these failure mechanisms can be connected to failure events that occur as a function of the applied strain. Here we will unify our microscopic insights with the macroscopic failure regimes described earlier based on the strain values at which these events occur. These results are summarized in Fig.~\ref{fig:MacroToMicro}. 

At the macroscopic level we identify two characteristic strains: the softening strain $\epsilon_{soft}$, where the stiffness ($d\sigma/d\epsilon$) drops below the initial stiffness (See Fig.~\ref{fig:StressStrain}(a)), and the failure strain $\epsilon_f$ where the system breaks in two pieces. At the microscopic level we identify strain points of three distinct events: the strain $\epsilon_{\Lambda,max}$ corresponding to the maximum in stress concentration (peak in Fig.~\ref{fig:StressConc}(e)), the strain where the diffuse damage is maximal $\epsilon_{diff,max}$, corresponding to the peak in Fig.~\ref{fig:Diffuse}(a), and finally the strain where the (geometrical) percolation of the network of sacrificial bonds is lost $\epsilon_{perc}$ (i.e. the top and bottom of the sacrificial network are no longer connected).

\emph{B regime -} In the brittle regime, far below the BDT, the system deforms homogeneously and every microcrack immediately develops into a macroscopic crack, leading to global failure. Therefore all failure events, both macroscopically and microscopically, occur at the failure strain $\epsilon_f$.

\emph{QB regime -} Approaching the BDT from below, we arrive at the QB regime, where global failure is preceded by softening in the stress-strain curve. The onset of this softening behaviour $\epsilon_{soft}$ becomes distinguishable from the failure strain $\epsilon_f$ around the transition from B-to-QB. Interestingly, the softening strain does not depend strongly on the distance from the BDT, $\Delta\tilde{\alpha}$ (Fig.~\ref{fig:MacroToMicro}). If we direct our attention to the microscopic level, we see that the onset of softening coincides with the peak in stress concentration $\Lambda$ (Fig.~\ref{fig:StressConc}(e)). Also for $\epsilon_{\Lambda,max}$ we only see a weak dependence on the distance from the BDT (Fig.~\ref{fig:MacroToMicro}). Here we would like to stress that even though the softening strain is the first clear macroscopic sign of failure, it is not the strain at which the first bonds break. Actually, a significant amount of (weak) bonds break homogeneously throughout the system before the softening strain. However, the creation of these microcracks does not have a big effect on the stress-strain response because the bonds surrounding these microcracks absorb the released stress, leading to the increase in stress concentration in Fig.~\ref{fig:StressConc}(e). At the softening strain, such a large amount of microcracks has been formed that the system can release (a part of) the stress caused by an applied strain via structural relaxation. i.e. the rearrangement of nodes and springs, causing the softening in the stress-strain curve in Fig.~\ref{fig:StressStrain}(a) and the drop in $\Lambda$ in Fig.~\ref{fig:StressConc}(e). Also after $\epsilon_{soft}$, we mostly observe the creation of new microcracks or a minor expansion of existing microcracks as indicated by the strong rise of diffuse failure (Fig.~\ref{fig:Diffuse}). However in the QB regime, just before final failure, most of these microcracks merge into one large macroscopic crack, leading to global failure as is visible in the middle snapshot in Fig.~\ref{fig:Diffuse}.

\emph{D1 regime -} After the BDT, the merging of microcracks starts to occur well before system failure, as can be identified from a peak at $\epsilon_{diff,max}$ in the diffuse failure as a function of strain (see Fig.~\ref{fig:Diffuse}(a)). In the ductile regime, the merging of cracks occurs at the same strain, as indicated by the fact that $\epsilon_{diff,max}$ is independent of the distance from the BDT. Even though microscopic cracks in the sacrificial network start to merge, this does not mean that the sacrificial network breaks completely, i.e. loses (geometrical) percolation. In fact, in the D1 regime for a system with sufficient disorder in $\lambda_S$ loss of percolation in the sacrificial network only occurs around the failure strain (Fig.~\ref{fig:MacroToMicro}). Thus, in a system with a network structure such as the RSN, the criterion for entering the plateau region in the stress-strain curve (Fig.~\ref{fig:StressStrain}) does not have to be loss of percolation in the sacrificial network, but rather the opportunity for structural relaxation within the sacrificial network. This observation provides an alternative mechanism with respect to what was proposed in literature, where loss of percolation was thought to be necessary to enter the ductile or necking regime~\cite{Sumiyoshi2006,Nakajima2013a}. 

\emph{D2 regime -} In the D2 regime, the stress-strain curve enters a matrix dominated regime after the plateau in stress (see Fig.~\ref{fig:StressStrain}(a)). Microscopically this transition is marked by the loss of percolation in the sacrificial network. From the loss of percolation onward, the stress in the material is carried by the matrix as is also apparent from the constant slope in the stress-strain curve (Fig.~\ref{fig:StressStrain}(a)). However, we do also observe that $\epsilon_f$ continues to grow with increasing $\Delta\tilde{\alpha}$, until a plateau is reached in the failure strain as we have observed earlier in Fig.~\ref{fig:FailureStrain}. The initial increase of $\epsilon_f$ in the D2 regime, shows that although percolation is lost in the sacrificial network and the matrix carries most of the stress, the sacrificial network can still contribute to the mechanical properties. Finally, we emphasize that during all these processes the matrix network remains largely intact and only very close to macroscopic failure the matrix bonds break (see Supplemental Material~\cite{supp}). 

\subsection{\label{ExperimentalConnection} Connecting the model to experiments}

In our model, we explored how stiffness ($\mu_M$ and $\mu_S$) and extensibility (strain at break of a single spring $\lambda_M$ and $\lambda_S$) of the individual networks influence the failure behaviour of double networks. Typically, in an experimental system, such as a hydrogel or an elastomer, those parameters cannot be tuned independently. For example, the stiffness is set by the density of polymer chains~\cite{James1943} and the extensibility is set by the length of these chains~\cite{Lake1967}. As a first approximation, the volume fraction of monomers determines the density of polymer chains, and thus the stiffness ($\mu$), whereas the ratio between the volume fraction of monomer and crosslinker determines the chain length, and thus the extensibility ($\lambda$). This is only an approximation, because at high monomer volume fractions both intra- and inter-network entanglements will occur. These entanglements can act as effective crosslinks, increasing the effective chain density and reducing the effective chain length \cite{Langley1974,Tsukeshiba2005}.

Nevertheless, within this approximate mapping, we find that experimental results for hydrogel and elastomer double networks are qualitatively consistent with the predictions of our model. For example, hydrogels show a transition from a brittle to a ductile failure response if the volume fraction of the matrix monomer is increased (i.e. an increase in $\mu_M$ in the model, consistent with Fig.~\ref{fig:PhaseDiagram})~\cite{Gong2003,Ahmed2014}. Moreover, experiments on hydrogels and elastomers reveal that an effective way to make a tough double network is to swell the sacrificial network~\cite{Ahmed2014,Nakajima2013,Matsuda2016,Millereau2018}, which pre-stretches the polymer chains in the sacrificial network. The pre-stretching reduces the extensibility of the sacrifical network, which can be captured in our model by decreasing $\lambda_S$. In agreement with experiments, we observe in Fig.~\ref{fig:PhaseDiagram} that by reducing $\lambda_S$ we can enter the ductile failure regime. If the matrix monomers are used to swell the network, both $\lambda_S$ and $\mu_M/\mu_S$ are affected, revealing that the experimental swelling protocol ultimately determines which regions of the phase diagrams in Fig. \ref{fig:PhaseDiagram} can be explored. Furthermore, our model shows that disorder tunes the macroscopic failure response away from the BDT, such as the B-to-QB transition and the D1-to-D2 transition. In experiments, the introduction of voids can be one method to introduce disorder in a controlled way. For example, in Ref.~\cite{Gong2011} a shift in the failure response is observed from D2 to D1 with an increase in void size. In the context of our model, this corresponds to an increase in disorder and can be explained as a decrease in $\Delta \tilde{\alpha}$ (Eq.~\ref{Datilde}). 

Finally, recent advances in mechanochemistry made it possible to explore the failure response at the microscopic level as well, revealing significant diffuse damage, even before the yielding point (Fig.~\ref{fig:StressConc}).~\cite{Ducrot2014,Millereau2018} In addition, scattering experiments show evidence of the delocalization of stress by probing the length scale of the non-affine response \cite{Ducrot2015} and the expansion of microscopic defects~\cite{Fukao2020}. In particular Ref.~\cite{Fukao2020} suggests that a higher fraction of matrix monomer leads to the creation of new microcracks instead of the expansion of existing defects (Fig.~\ref{fig:Diffuse}). These observations fit within our theoretical framework where the distance from the BDT and disorder provide control over the microscopic failure mechanism.

The current simulation framework can be extended to increase the accuracy of its predictions for polymeric systems such as hydrogels and elastomers. In particular, we expect that the implementation of non-linear springs will better represent the non-linear stress-strain response of finitely extensible polymers and the redistribution of stress in a polymer network. With such an extended model we could explore if non-linear elasticity enhances stress delocalization with respect to linear elasticity and if this effect influences nucleation and propagation of microcracks.

\section{\label{sec:SummConclusions}Summary and Conclusions}
In this paper we studied both the macroscopic and microscopic failure behavior of double network materials using a spring network model (RSN) with emergent load sharing. By comparing the results of the RSN with a model based on equal load sharing, we reveal that: (i) The location of the BDT, defined on the basis of macroscopic stress-strain behavior, is captured by a simple force balance.
(ii) Disorder introduces intermediate failure regimes but it can be incorporated in the parameter $\Delta\tilde{\alpha}$ to correctly describe the distance from the BDT, allowing for rescaling of the number of total broken sacrificial bonds. (iii) At the microscopic level stress concentration and delocalization reveal a markedly different picture compared to global load sharing. 

The overall picture that emerges from the RSN is that the force balance, a central feature of double networks, has significant control over both the macroscopic and microscopic failure behaviour, irrespective of how stress is (re)distributed. By contrast, the nucleation and propagation of (micro)cracks is also highly dependent on the mode of stress (re)distribution. In particular we have identified how stress concentration, diffuse damage and loss of percolation are related to the transitions from B-to-QB, QB-to-D1, and D1-to-D2, respectively.

We highlight that, because many microcracks can form before global failure due to the stabilization by the matrix, the load sharing in double networks becomes highly non-linear as a result of the interaction between these microcracks. Therefore, double networks provide a unique opportunity to exploit these non-linearities in microcrack interaction. For example, this knowledge about the microscopic failure process could aid the development of robust self-healing double networks, as the healing of diffuse damage is easier than the healing of macroscopic cracks~\cite{Jia2016,Liu2017,Yan2017}.

By extending the RSN model, additional features of double network gel failure could be studied, such as the influence of pre-stress or structural disorder in the sacrificial network. Furthermore the introduction of disorder in the matrix would allow to also study the enhancement of strength and toughness, including the role of microscopic failure. In conclusion, this paper demonstrates that a random spring network model provides the opportunity to systematically study the microscopic failure process within double networks.

\section*{Acknowledgements}
The work of J.T., S.D. and J.v.d.G. is part of the SOFTBREAK project funded by the European Research Council (ERC Consolidator Grant).

\bibliographystyle{unsrt}
\end{document}


\maketitle

\section{\label{sec:failureRegimesAnalysis}Classification of stress-strain responses}
\begin{figure*}
    \centering
    \includegraphics{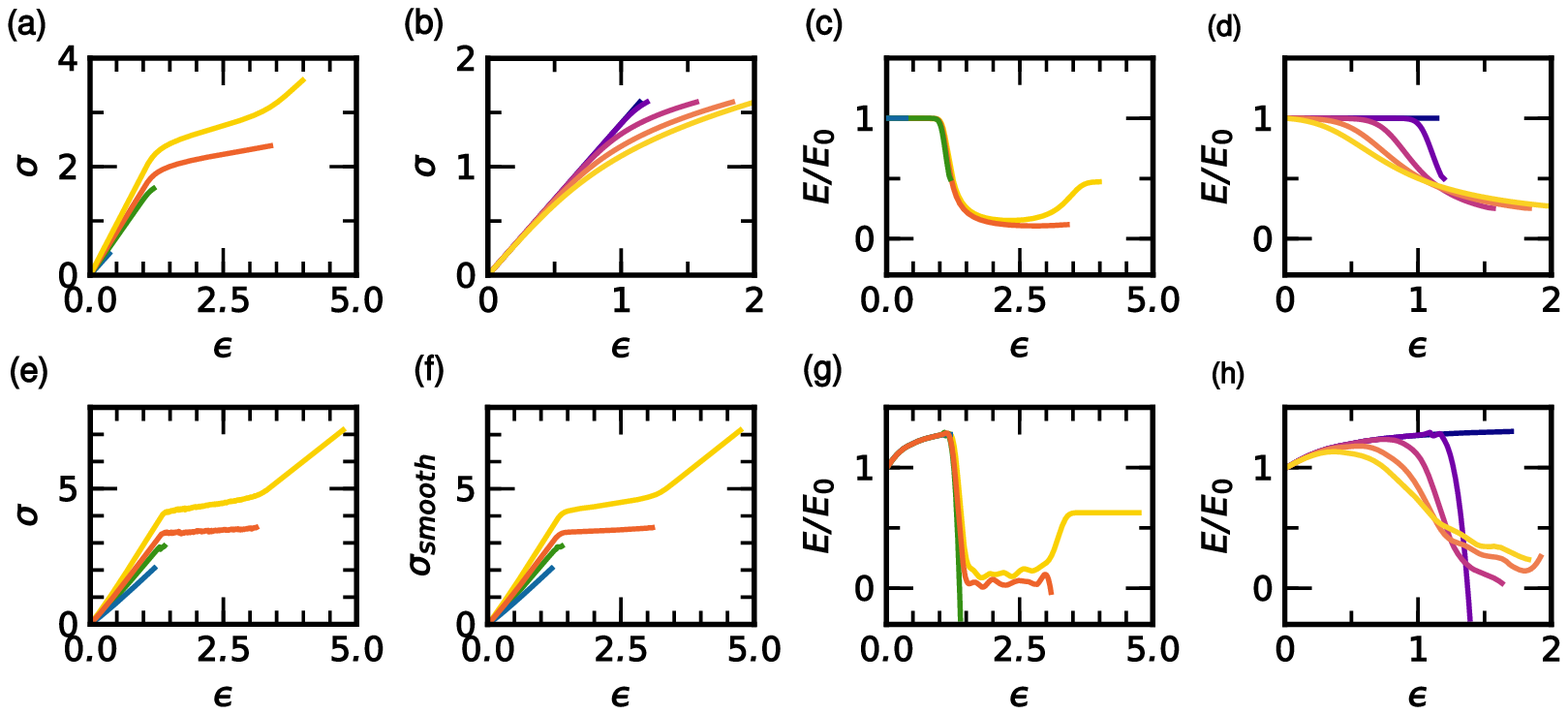}
    \caption{Stress-strain classification based on stiffness for both the MS model ((a)-(d)) and the RSN model ((e)-(h)). For the MS model we show in panel (a) the stress-strain curves for stiffness ratios $\mu_M/\mu_S$ $0.1$ (blue),$0.4$ (green)),$0.6$ (orange) or $0.8$ (yellow) ($\left<\lambda_S\right>=1.4$,$\delta\lambda=0.125$). (b) the stress-strain curves for different amounts of bond strength disorder $\delta\lambda$ from $0.000$ (blue) to $0.500$ (yellow) ($\left<\lambda_S\right>=1.4$,$\mu_M/\mu_S=0.4$). panels (c) and (d) show the stiffness corresponding to panels (a) and (b) normalized by the initial stiffness $E_0$. For the RSN model we show in panel (e) stress-strain curves for a series of stiffness ratios (see description of panel (a) for details). (f) the same stress-strain curves smoothed according to the description in section \ref{sec:failureRegimesAnalysis}. (g) the normalized stiffness corresponding to the the stress-strain curves in panel(f). (h) the normalized stiffness of for different amounts of disorder (see description of panel (b) for details).}
    \label{fig:ManySpringMacroscopicResponse}
\end{figure*}

We use the following procedure to classify the stress-strain response of the different networks. We define four categories: brittle (B), quasibrittle (QB) and two ductile phases (D1 and D2). The characterization is based on the stiffness of the network during the deformation, defined as the derivative of the stress-strain response, normalized by the initial stiffness. Examples are shown in Supp. Fig. \ref{fig:ManySpringMacroscopicResponse}. If a material is brittle (B), its stiffness during extension will stay constant, or at least will not drop far below its initial stiffness. In case of quasibrittle (QB) failure, softening occurs, which causes a sharp drop in material stiffness. We identify the softening strain as the point where $E/E_0$ drops below $1.0$, the initial stiffness. A material that shows ductile failure also goes through this softening stage, but after softening a plateau can be seen in the stiffness curve.  The transition from QB to D1 is based on the existence of the plateau. The plateau is identified as the strain region where $E(\epsilon)/E_0\leq E_{min}/E_0 + 0.05$, provided $E_{min}/E_0<0.7$. This region must span at least 50\% of strain in order to be identified as a plateau, otherwise we classify the failure response as QB. If the material breaks in this plateau region, it is classified as D1. If instead the material response exceeds this plateau and achieves a higher stiffness, we classify this response as D2. We note that in our case the final stiffness is always lower than the initial stiffness, since the last part of the D2 response is controlled by the matrix elasticity for which we have $\mu_M/\mu_S < 1.0$. This final strain region is identified as the region from the end of the plateau until final failure. The stiffness in this plateau must exceed $E_0$ by 15\% in order to be identified as D2.

For the MS model the stiffness is determined as the linear piece-wise derivative of the stress-strain curve. For the RSN model, the stress-strain response is smoothed before determining the failure response in order to reduce the influence of small fluctuations due to the failure of bonds. The smoothing procedure has 4 steps. Firstly only strain points are selected that have a higher stress than the previous points. Secondly, we make the data series equispaced in strain via linear interpolation with an interval of $0.1\%$ strain. Thirdly, we smooth the curve via a linear fit over a window of $2\%$ strain, the resulting stress-strain curve has a strain interval of $1\%$ strain. Finally we determine the stiffness (first derivative of the stress-strain curve) using a Savitsky-Golay algorithm (4th order and $41\%$ strain window). At the edges half of the window size is fitted with a 4th order polynomial.

\section{Crack detection algorithm}
To monitor the microscopic failure patterns in the RSN model, we keep track of the formation and development of cracks in the sacrificial network. For simplicity, in this analysis we only consider sacrificial bonds that align with the strain field and therefore carry most of the stress (in our case, these are the non-horizontal initial bonds). At the beginning of the simulation, the sacrificial network is intact but as soon as bonds break voids appear. These voids are defined as cracks. By studying the number of voids and their size in terms of the number of broken sacrificial bonds, we obtain a complete picture of the microscopic failure behavior. In particular, we calculate the size of the largest crack with respect to the total number of broken sacrificial bonds and express this as the fraction $\phi_{large}$. All the broken sacrificial bonds that do not end up in the largest crack represent the non-localized damage occurring throughout the network. We define $\phi_{diff}=\phi-\phi_{large}$ to quantify the diffuse damage.

\begin{figure*}
    \centering
    \includegraphics{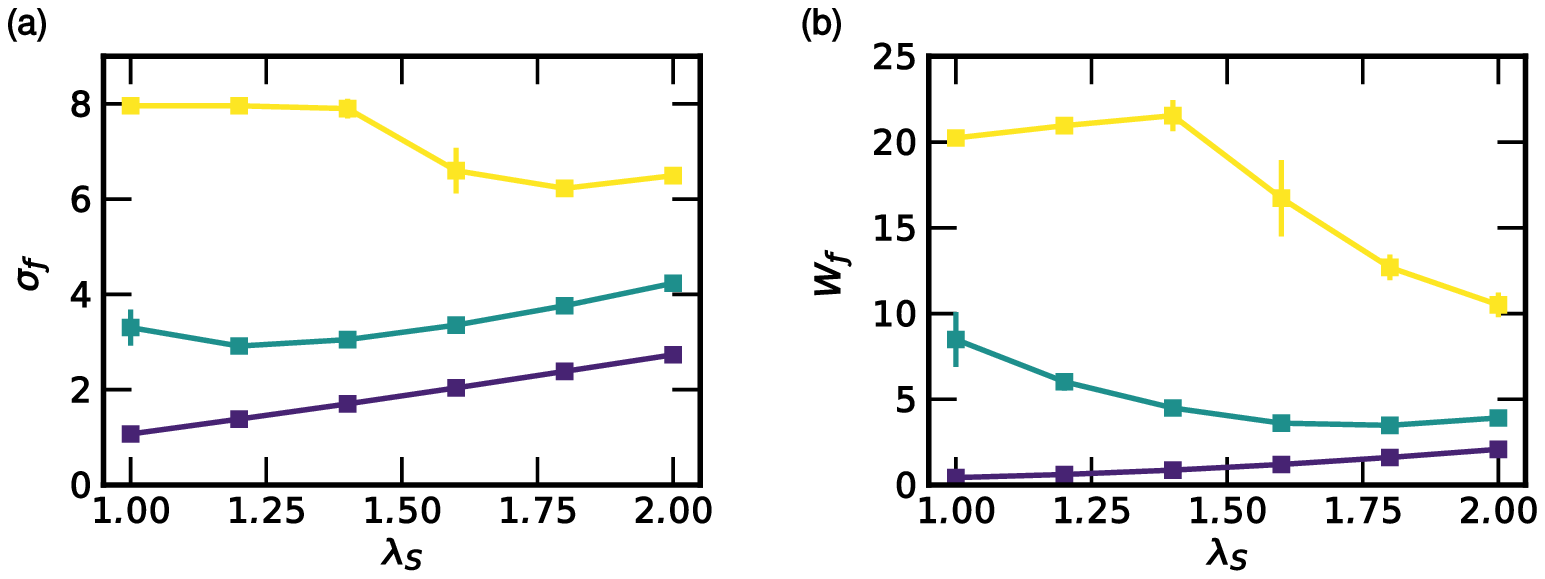}
    \caption{ Material properties versus the average failure threshold $\langle \lambda_S \rangle$. ($\delta\lambda=0.250$, $L=50$). (a) Failure stress, $\sigma_f$. (b) Work of extension $W_f$, the area under the stress-strain curve untill global failure. Results are shown for stiffness ratios $\mu_M/\mu_S =$ 0.10 (purple), 0.50 (green) and 1.00 (yellow)).}
    \label{fig:FailureStressStrain}
\end{figure*}
 
\begin{figure*}
    \centering
    \includegraphics{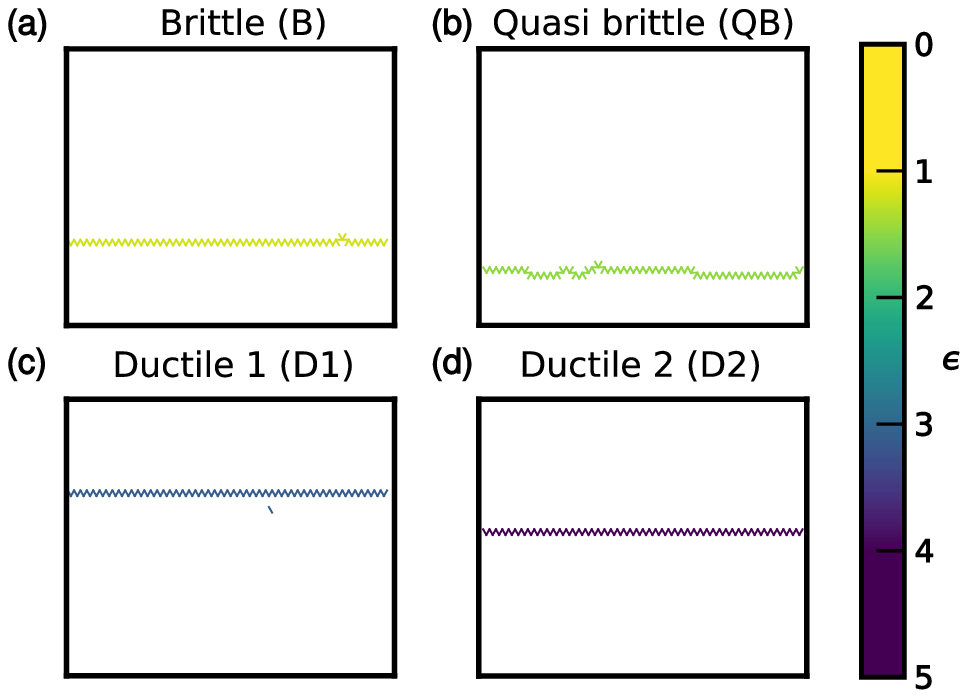}
    \caption{Spatial failure in the matrix network for the RSN model. (a)-(d) Failure patterns in the matrix network for the four distinct failure behaviours. Every line represents a broken matrix bond, color-coded according to the strain at which the bond fails (see color bar above).}
    \label{fig:MatrixNetworkFailure}
\end{figure*}

\begin{figure*}
    \centering
    \includegraphics{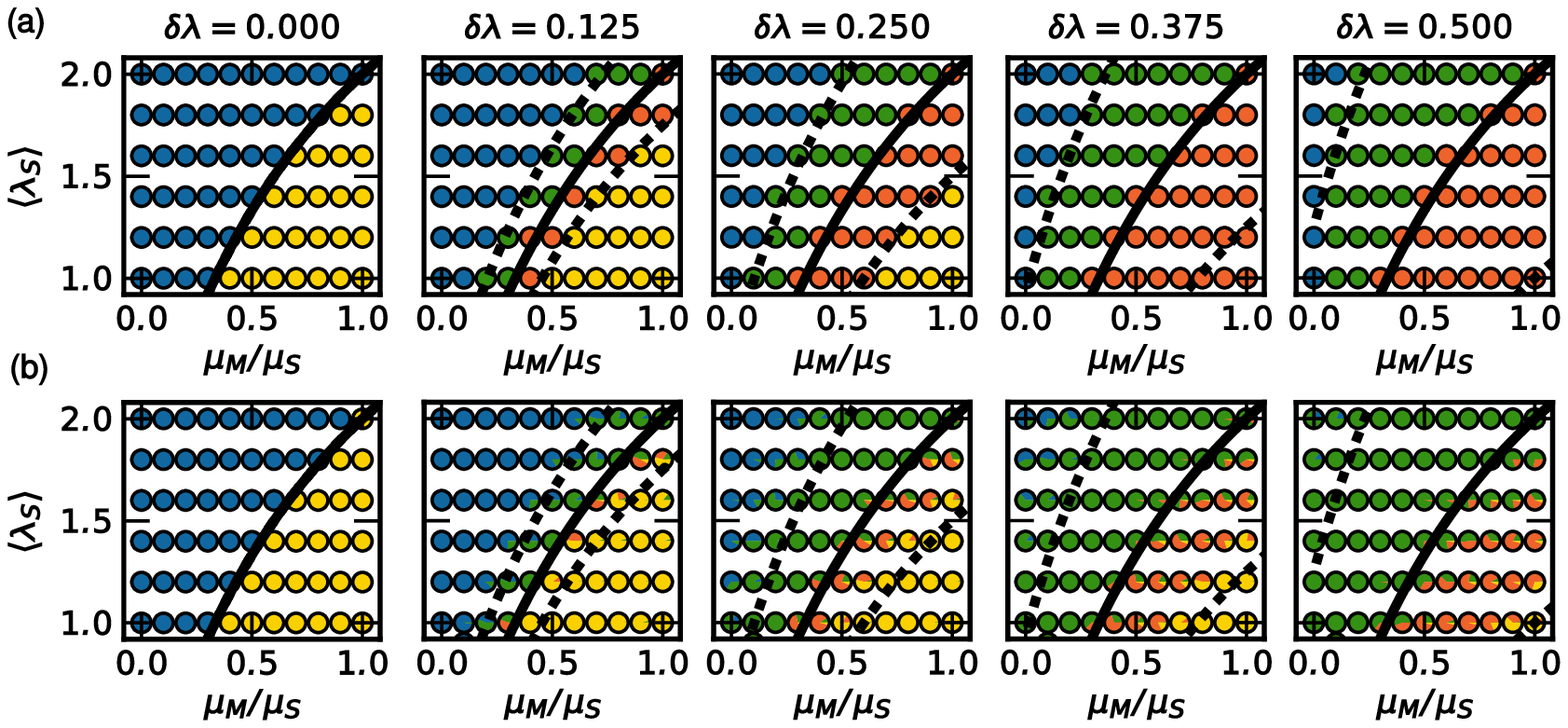}
    \caption{ Stress-strain response of both the MS model (panel (a)) and the RSN model (panel (b)) in the ($\mu_M/\mu_S,\lambda_S$)-plane for different amounts of disorder $\delta\lambda$. See \ref{sec:failureRegimesAnalysis} for more information. The colors correspond to the failure regimes: brittle (blue), quasi-brittle (green), ductile 1 (orange) and ductile 2 (yellow). The black line indicates the BDT transition as predicted by the MS model, the dotted lines are placed at $\Delta\tilde{\alpha}=-2.0$ and $\Delta\tilde{\alpha}=2.5$ corresponding to the estimated B-to-QB and D1-to-D2 transitions in the MS model respectively.}
    \label{fig:ManySpringsDiagramStressStrain}
\end{figure*}

\begin{figure*}
    \centering
    \includegraphics{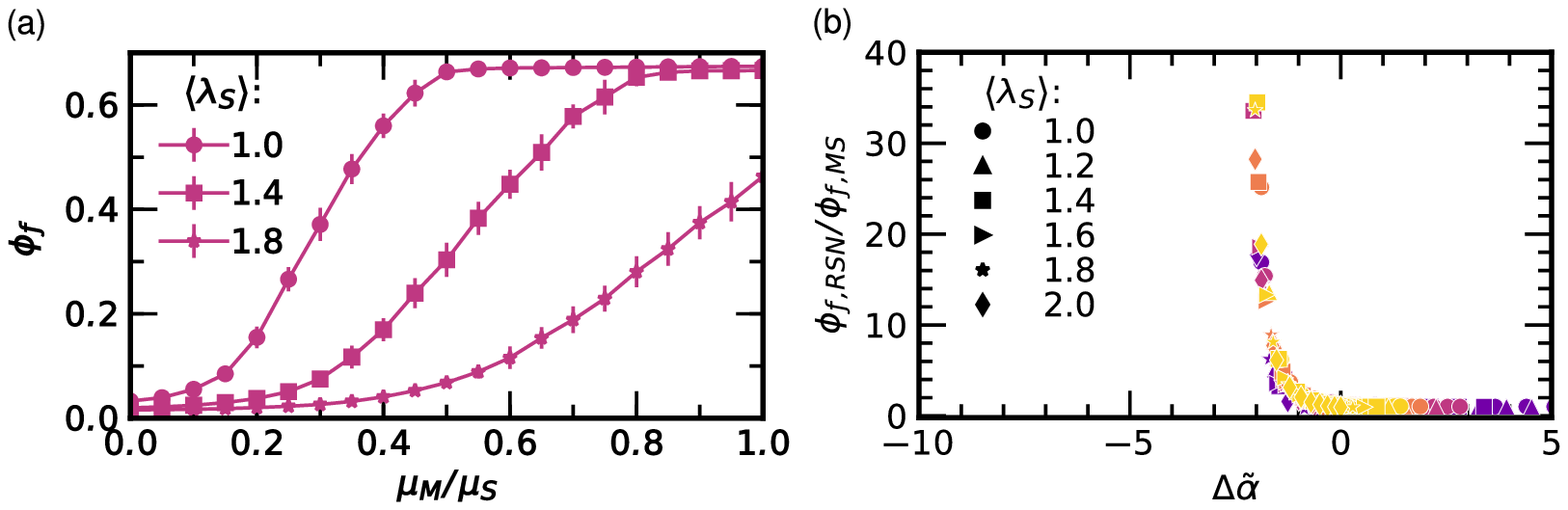}
    \caption{ (a) Fraction of broken sacrificial bonds in the RSN model after macroscopic failure $\phi_f$ as a function of $\mu_M/\mu_S$. Several $\langle\lambda_S\rangle$. $\delta\lambda=0.250$ and $L=50$. (b) The relative increase in $\phi_{f,RSN}$ with respect to $\phi_{f,MS}$ plotted versus $\Delta\tilde{\alpha}$ for a range of $\lambda_S$ (marker shape) and a range of $\delta\lambda$ (marker color).)}
    \label{fig:UnscaledBrokenbonds}
\end{figure*}

\begin{figure*}
    \centering
    \includegraphics{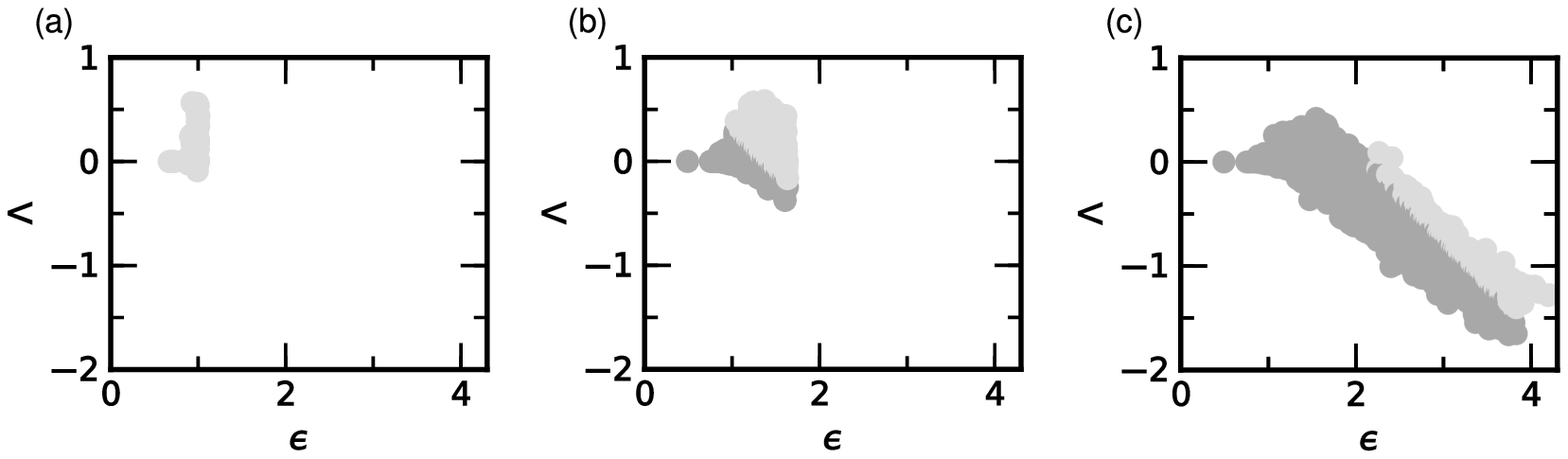}
    \caption{Failure of weak and strong bonds. For a single simulation $\Lambda=\lambda_{S,fail}-dl_{aff}$ is plotted versus $\epsilon$. Every dot represents a broken sacrificial bond and the color indicates if the bond is strong (light gray) or weak (dark gray). In all panels $\langle\lambda_S\rangle=1.4$ and $\delta\lambda=0.250$. (a) $\mu_M/\mu_S=0.1$. (b) $\mu_M/\mu_S=0.4$. (c) $\mu_M/\mu_S=0.8$. }
    \label{fig:StressConcentrationWeakStrong}
\end{figure*}